\documentclass[journal=jpccck,manuscript=article]{achemso}
\setkeys{acs}{usetitle=true,maxauthors=99, doi=true}
\pdfoutput=1
\usepackage{doi}
\usepackage{hyperref}
\AtBeginDocument
  {%
    
  }
\usepackage[version=3]{mhchem} 

\usepackage{graphicx}
\usepackage{bm}

\usepackage[utf8]{inputenc}
\usepackage[T1]{fontenc}
\usepackage{mathtools}
\usepackage{amsmath, amssymb}

\author{Nicolas G. H{\"o}rmann}%
 \email{hoermann@fhi.mpg.de}
 \affiliation{Fritz-Haber-Institut der Max-Planck-Gesellschaft, Faradayweg 4-6, 14195 Berlin, Germany.
}%

\author{Simeon D. Beinlich}%
\affiliation{Fritz-Haber-Institut der Max-Planck-Gesellschaft, Faradayweg 4-6, 14195 Berlin, Germany.
}%
\alsoaffiliation{Technical University of Munich, Lichtenbergstr. 4, 85747 Garching, Germany}

\author{Karsten Reuter}%
 \affiliation{Fritz-Haber-Institut der Max-Planck-Gesellschaft, Faradayweg 4-6, 14195 Berlin, Germany.
}%

\title{Converging Divergent Paths: Constant Charge vs. Constant Potential Energetics in Computational Electrochemistry}

\begin{document}



\begin{abstract}
Using the example of a proton adsorption process, we analyze and compare two prominent modelling approaches in computational electrochemistry at metallic electrodes — electronically canonical, constant-charge and electronically grand-canonical, constant-potential calculations.
We first confirm that both methodologies yield consistent results for the differential free energy change in the infinite cell size limit. This validation emphasizes that, fundamentally, both methods are equally valid and precise. In practice, the grand-canonical, constant-potential approach shows superior interpretability and size convergence as it aligns closer to experimental ensembles and exhibits smaller finite-size effects. On the other hand, constant-charge calculations exhibit greater resilience against discrepancies, such as deviations in interfacial capacitance and absolute potential alignment, as their results inherently only depend on the surface charge, and not on the modelled charge vs. potential relation. The present analysis thus offers valuable insights and guidance for selecting the most appropriate ensemble when addressing diverse electrochemical challenges.
\end{abstract}

\maketitle

\section{Introduction}

Electrochemical processes at electrified interfaces play a pivotal role in a plethora of applications, ranging from energy storage systems to catalysis. Assessing the nuances of such processes by means of first-principles-based, atomistic simulations has long been a topic of interest for researchers. Corresponding state-of-the-art studies presently fall into two major classes:\cite{Abidi2021Atomistic} There is on the one hand the canonical, constant-charge approach\cite{Noerskov2005Trends,Noerskov2004Origin,Noerskov2009Towards, Hoermann2015Some,Skulason2007Density,rossmeisl2008modeling,Skulason2010Modeling,Chan2015Electrochemical} -- as popularized by the computational hydrogen electrode (CHE) method\cite{Noerskov2005Trends,Noerskov2004Origin,Noerskov2009Towards} -- and on the other hand the constant-potential approach, that treats the electronic degrees of freedom grand canonically and thus at constant-potential boundary conditions.\cite{Taylor2006First,bonnet2012first,goodpaster2016identification, sundararaman2017jdftx, sundararaman2017grand, Bouzid2017Redox,kastlunger2018controlled, Huang2018Potential, Hoermann2019Grand, melander2018grand, Lindgren2020Challenge, Hoermann2020Electrosorption,hagiwara2021bias, xia2023grand, surendralal2018first, deissenbeck2021dielectric, deissenbeck2023dielectric}
While both methods have been used extensively, a comprehensive understanding of their equivalences, practical differences, and the implications of these differences remains an area where clarity is sought.\cite{Gauthier2019Unified,Hoermann2020Electrosorption,Abidi2022How,beinlich2022field,dominguezflores2023approximating,beinlich2023controlled}

Using the adsorption of a proton at metallic electrodes as a prototypical electrochemical modelling problem, this study assesses the fundamental properties and practical implications of both computational approaches. First we prove that in the thermodynamic limit, both methods offer identical results concerning the differential free energy change. Subsequently, we elaborate in more detail on the advantages of the grand-canonical method in terms of interpretability and size convergence, but also their inferior sensitivity to robustness against certain modelling parameter errors. Consequently, this work offers vital guidelines in terms of validity, accuracy, and applicability of both methods and can aid researchers in selecting the most suitable ensemble for their specific electrochemical modelling requirements.

\section{Theory}
\subsection{Canonical, Constant-Charge Description of Proton Adsorption}

Within the canonical constant-charge (CC) approach by N{\o}rskov \emph{et al.} \cite{Noerskov2005Trends,Noerskov2004Origin} a proton adsorption reaction is assessed by evaluating the free energy change 
\begin{equation}
G^{\mathrm{s}}(N_{\mathrm{H^+}}+1, N_{\mathrm{e^-}}+1) - G^{\mathrm{s}}(N_{\mathrm{H^+}}, N_{\mathrm{e^-}}) \label{eq:finitediffCC}
\end{equation}
of an explicitly modelled electrode surface s while adding a proton-electron pair (effectively a hydrogen atom) consistent with the reaction equation 
\begin{equation}
    \mathrm{H^+} + e^- + *\rightarrow  {^*\mathrm H}  \label{eq:reactionHads} \quad .
\end{equation}
The method thus approximates the derivative $\frac{\mathrm{d} G^{\mathrm{s}}}{\mathrm{d}  N_{\mathrm{H}}}$ and maps the obained results to a free energy change $\Delta G$ at electrochemical conditions by including the energy cost to remove the exchanged particles from their thermodynamic baths according to
\begin{eqnarray}
\Delta G&=& \frac{\mathrm{d} G^{\mathrm{s}}}{\mathrm{d}  N_{\mathrm{H}}} - (\tilde{\mu}_{\mathrm{H^+}} + \tilde{\mu}_{e^-})\label{eq:CHE} \ . \label{eq:Can_DeltaG}
\end{eqnarray}
$\tilde{\mu}_{\mathrm{H^+}}$ hereby denotes the electrochemical potential of the proton and $\tilde{\mu}_{e^-}$ the electrochemical potential of the electron, which is trivially related to the electrode potential $\phi_E$ via $\tilde{\mu}_{e^-}= -e\phi_E$. The potential dependence $\propto\tilde{\mu}_{e^-}$ enters thus only in the {\em a posteriori} energetic referencing step, eq. (\ref{eq:CHE}), and not at the electronic structure calculation level, i.e., when evaluating $G^{\mathrm{s}}$. Referencing is typically done against the equilibrium at SHE conditions where $\frac{1}{2}{\mathrm{H_2(g)}} \leftrightharpoons {\mathrm{H^+(aq)}} + e^{-}\mathrm{(m)}$ and thus $ \tilde{\mu}^{\mathrm{SHE}}_{\mathrm{H^+}} + \tilde{\mu}^{\mathrm{SHE}}_{e^{-}} = \frac{1}{2} \mu_{\mathrm{H_2(g)}}$. It allows thereby straightforward referencing of $G^{\mathrm{s}}$ against $\mu_{\mathrm{H_2(g)}}$ within the so-termed CHE method.

Conventionally, one approximates $\frac{\mathrm{d} G^{\mathrm{s}}}{\mathrm{d}  N_{\mathrm{H}}} \sim \left.\frac{\mathrm{d} G^{\mathrm{s}}}{\mathrm{d}  N_{\mathrm{H}}}\right|_{\mathrm{PZC}}$ from the energy difference of DFT total energy calculations whose composition differs by exactly one $\mathrm{H}$ atom ($=\mathrm{H}^+ + e^-$) under potential of zero charge (PZC) conditions ($\tilde{\mu}^{\mathrm{PZC}}_{e^-}$), i.e., in supercell calculations without electronic excess charges. If $\Delta G$ shows a more complex behavior, e.g. a dependence on pH, potential, or electrolyte composition one is bound to make $\frac{\mathrm{d} G^{\mathrm{s}}}{\mathrm{d}  N_{\mathrm{H}}}$ a function of the respective variables. The CHE framework then posits that such effects can be computed by setting up the appropriate interface conditions, e.g. with included interfacial electrostatic fields $\mathbf{E}$ or electronic surface excess charges $-eN^{\mathrm{net}}_{e^-}$. Such $\mathbf{E}$-field-augmented\cite{Pasumarthi2023Comparative} CHE calculations then evaluate canonical energy differences at constant, finite fields\cite{Koper2000Field,Noerskov2004Origin,kelly2020electric} $\left.\frac{\mathrm{d} G^{\mathrm{s}}}{\mathrm{d}  N_{\mathrm{H}}}\right|_{\mathbf{E}}$ or similarly at a constant number of excess electrons\cite{Ringe2019Understanding, Gauthier2019Unified} $\left.\frac{\mathrm{d} G^{\mathrm{s}}}{\mathrm{d}  N_{\mathrm{H}}}\right|_{N^{\mathrm{net}}_{e^-}}$. Both methods add essentially dipole-field-like energy contributions (see discussion below and section S1 in the SI) and thus yield by and large identical results if $N^{\mathrm{net}}_{e^-}$ is chosen consistent with the field $\mathbf{E}$.\cite{Dudzinski2023First} While the so-discussed CHE methods are thus based on canonical energy differences of systems of predefined charge states (e.g. fixed electronic excess charge $N^{\mathrm{net}}_{e^-}$), more recent, constant potential methods approach the proton adsorption process at electrified interfaces differently, by modelling directly the effect of the potential.

\subsection{Grand Canonical Description of Proton Adsorption}

Constant-potential descriptions treat the electronic degree of freedom grand canonically, i.e., they allow the electron number to adapt according to the constant-potential boundary condition. The intellectual difference to the CHE method becomes clearer when we reanalyze the free energy change for the adsorption reaction, eq. (\ref{eq:reactionHads}). Within the constant-potential approach\cite{Hoermann2019Grand,Hoermann2020Electrosorption,Ge2020Coupling,Lindgren2020Challenge,Ringe2022Implicit} energy differences are to be taken for energies $G^{\mathrm{s}}$ that were Legendre-transformed in the electronic degrees of freedom in order to yield the electronically grand-canonical (GC) energies $\mathcal{G}^{\mathrm{s}}$ at given electron electrochemical potential $\tilde{\mu}_{e^-}$, where
\begin{eqnarray}
\mathcal{G}^{\mathrm{s}}(N_{\mathrm{H^+}}, \tilde{\mu}_{e^-}) &=& 
G^{\mathrm{s}}(N_{\mathrm{H^+}}, N_{e^-})
- N_{e^-}\tilde{\mu}_{e^-} \ . \label{eq:LT} 
\end{eqnarray}
Here, it is of central importance to recognize that $N_{e^-} = N_{e^-}(N_{\mathrm{H^+}},\tilde{\mu}_{e^-})$ is a function of the thermodynamic variables $N_{\mathrm{H^+}}$ and $\tilde{\mu}_{e^-}$. The GC energy change of the interfacial system when adding a proton is then obtained as the finite difference
\begin{equation}
\mathcal{G}^{\mathrm{s}}(N_{\mathrm{H^+}}+1, \tilde{\mu}_{e^-}) - \mathcal{G}^{\mathrm{s}}(N_{\mathrm{H^+}}, \tilde{\mu}_{e^-}) \quad ,
\end{equation}
which approximates $\left.\frac{\partial \mathcal{G}^{\mathrm{s}}}{\partial  N_{\mathrm{H^+}}}\right|_{\tilde{\mu}_{e^-}}$ at fixed electrode potential and thus allows to evaluate the energy change of the proton adsorption reaction as 
\begin{eqnarray}
\Delta \mathcal{G} &=& \left.\frac{\partial \mathcal{G}^{\mathrm{s}}}{\partial  N_{\mathrm{H^+}}}\right|_{\tilde{\mu}_{e^-}} - \tilde{\mu}_{{\mathrm{H^+}}} \label{eq:GCdeltaG} \quad .
\end{eqnarray}
Note that GC simulations thus simply and exclusively describe a proton adsorption process. The concomitant number of exchanged electrons is only obtained {\em a posteriori} from the simulations at fixed ${\tilde{\mu}_{e^-}}=\mathrm{const}$, and GC calculations are valid and applicable independent of charge-transfer considerations\cite{Lindgren2020Challenge}. This is in contrast to the CHE method which assesses the canonical energy difference by a simultaneous exchange of a proton-electron pair $\mathrm{H}^+ + e^-$ as expressed by the derivative in the number of \emph{hydrogen atoms} ${\mathrm{H}}$ as in eq.~(\ref{eq:Can_DeltaG}).

\section{Comparative Analysis}
\subsection{Relation between both Descriptions}
\label{sec:PES}

In order to understand the relation between $\Delta G$ and $\Delta \mathcal{G}$ one needs to analyze how the partial derivative $\left.\frac{\partial \mathcal{G}^{\mathrm{s}}}{\partial  N_{\mathrm{H^+}}}\right|_{\tilde{\mu}_{e^-}}$ can be reexpressed in terms of derivatives of the canonical Gibbs energy 
$G^{\mathrm{s}}(N_{\mathrm{H^+}}, N_{e^-})$. Using the Legendre transform, eq. (\ref{eq:LT}), one can write 
\begin{eqnarray}
\left.\frac{\partial \mathcal{G}^{\mathrm{s}}}{\partial  N_{\mathrm{H^+}}}\right|_{\tilde{\mu}_{e^-}} = \left.\frac{\mathrm{d} G^{\mathrm{s}}}{\mathrm{d} N_{\mathrm{H^+}}}\right|_{\tilde{\mu}_{e^-}} - 
\left.\frac{\partial N_{e^-}}{\partial  N_{\mathrm{H^+}}}\right|_{\tilde{\mu}_{e^-}} \tilde{\mu}_{e^-} \quad , \label{eq:deriv1}
\end{eqnarray}
where the total derivative (first term) signifies that one has to consider the total change of $G^{\mathrm{s}}(N_{\mathrm{H^+}},N_{e^-})$ at constant $\tilde{\mu}_{e^-}$ which arises on the one hand from the change of $N_{\mathrm{H^+}}$ but as well from the adaption in $N_{e^-} = N_{e^-}(N_{\mathrm{H^+}},\tilde{\mu}_{e^-})$ at constant potential conditions. The second term represents the energy cost to take out the respective total number of electrons from the bath which derives from the second term in the Legendre transform in eq.~(\ref{eq:LT}).

Without loss of generality, the change in $N_{e^-}$ at metallic electrodes can be decomposed into
\begin{eqnarray}
\left.\frac{\partial N_{e^-}}{\partial  N_{\mathrm{H^+}}}\right|_{\tilde{\mu}_{e^-}}&=& 1 + \left.\frac{\partial N^{\mathrm{net}}_{e^-}}{\partial  N_{\mathrm{H^+}}}\right|_{\tilde{\mu}_{e^-}} \ , \label{eq:esvla}
\end{eqnarray}
where the 1 arises from the exchange of one electron to compensate for the +1 charge of the proton at the interface, and the second term on the right hand side derives from the consecutive adaption of the double layer charge $N^{\mathrm{net}}_{e^-}$ to keep the electrode potential fixed. In the same logic, one can decompose the first term in eq.~(\ref{eq:deriv1}), $\left.\frac{\mathrm{d} G^{\mathrm{s}}}{\mathrm{d} N_{\mathrm{H^+}}}\right|_{\tilde{\mu}_{e^-}}$, into a first part at constant double layer charge ($N^{\mathrm{net}}_{e^-} = \mathrm{const.}$) and a second part that accounts for the change in $N^{\mathrm{net}}_{e^-}$. Evidently, as we exchange identically one proton and one electron in the former contribution, it is identical to the considered energy difference in the CHE method, and will thus be written for convenience (and consistent with the considerations above) as the total derivative in the number of hydrogen atoms (see section S2 in the SI). We thus obtain
\begin{eqnarray}
\left.\frac{\mathrm{d} G^{\mathrm{s}}}{\mathrm{d} N_{\mathrm{H^+}}}\right|_{\tilde{\mu}_{e^-}} &=& \left.\frac{\mathrm{d} G^{\mathrm{s}}}{\mathrm{d} N_{\mathrm{H}}}\right|_{N^{\mathrm{net}}_{e^-}({\tilde{\mu}_{e^-}})} + \tilde{\mu}_{e^-} \left.\frac{\partial N^{\mathrm{net}}_{e^-}}{\partial  N_{\mathrm{H^+}}}\right|_{\tilde{\mu}_{e^-}} \quad , \label{eq:eq2}
\end{eqnarray}
from the identity $\frac{\partial G^{\mathrm{s}}}{\partial  N_{e^-}} = \tilde{\mu}_{e^-}$ at equilibrium. Insertion of eqs.~(\ref{eq:esvla}) and (\ref{eq:eq2}) into eq.~(\ref{eq:deriv1}) then yields
\begin{eqnarray}
\left.\frac{\partial \mathcal{G}^{\mathrm{s}}}{\partial  N_{\mathrm{H^+}}}\right|_{\tilde{\mu}_{e^-}} &=& \left.\frac{\mathrm{d} G^{\mathrm{s}}}{\mathrm{d} N_{\mathrm{H}}}\right|_{N^{\mathrm{net}}_{e^-}({\tilde{\mu}_{e^-}})} - \tilde{\mu}_{e^-} \label{eq:beforelast} \quad .
\end{eqnarray}
By inserting this back into eq.~(\ref{eq:GCdeltaG}) and comparison with eq.~(\ref{eq:Can_DeltaG}) we then find $\Delta G = \Delta \mathcal{G}$ when canonical energy differences $\left.\frac{\mathrm{d} G^{\mathrm{s}}}{\mathrm{d} N_{\mathrm{H}}}\right|_{N^{\mathrm{net}}_{e^-}({\tilde{\mu}_{e^-}})}$ at constant double-layer charge are used and whenever the relation ${N^{\mathrm{net}}_{e^-}({\tilde{\mu}_{e^-}})}$ is identical in the canonical and grand-canonical treatment. Hence, for metallic electrodes, GC methods are expected to yield identical results with the ($\mathbf{E}$-field-augmented) CC CHE method in the thermodynamic limit, i.e., whenever energy differences become differentials. Importantly though, we stress that this applies only if the proton state under study is identical and well defined (stable) in both ensembles. As we showed recently, one can in principle stabilize states in the canonical ensemble that are not stable against infinitesimal geometric perturbations and thus not observable in the GC ensemble \cite{beinlich2023theoretical}.

\subsection{Advantages and Disadvantages of Both Methods}
While thus $\Delta \mathcal{G}=\Delta G$ numerically for calculations in the infinite cell size limit, both methods have advantages and disadvantages in terms of interpretability, size convergence and sensitivity to computational errors. 
\paragraph{Interpretability}
The meaning of $\Delta \mathcal{G}$ is arguably more evident within the GC picture: Considering eq.~(\ref{eq:GCdeltaG}), $\Delta \mathcal{G}$ simply measures the difference between $\left.\frac{\partial \mathcal{G}^{\mathrm{s}}}{\partial  N_{\mathrm{H^+}}}\right|_{\tilde{\mu}_{e^-}}$ and the proton electrochemical potential in the bulk solution reservoir $\tilde{\mu}_{{\mathrm{H^+}}}$.  
Note that one can rewrite this partial derivative as
\begin{eqnarray}
\left.\frac{\partial \mathcal{G}^{\mathrm{s}}}{\partial  N_{\mathrm{H^+}}}\right|_{\tilde{\mu}_{e^-}} &=&
\left.\frac{\partial G^{\mathrm{s}}}{\partial  N_{\mathrm{H^+}}}\right|_{N_{e^-}(\tilde{\mu}_{e^-})} \label{eq:equiv_gc-c} \\
&+& \underbrace{ \frac{\partial G^{\mathrm{s}}}{\partial  N_{e^-}} \frac{\partial N_{e^-}}{\partial  N_{\mathrm{H^+}}} - \frac{\partial N_{e^-}}{\partial  N_{\mathrm{H^+}}} \tilde{\mu}_{e^-}}_{ = 0} \ , \nonumber
\end{eqnarray}
using eq.~(\ref{eq:LT}) and the identity $\frac{\partial G^{\mathrm{s}}}{\partial  N_{e^-}} = \tilde{\mu}_{e^-}$ at equilibrium. The right hand side of eq.~(\ref{eq:equiv_gc-c}) is the standard expression of the electrochemical potential of the proton as it is the partial derivative of the canonical energy $G^{\mathrm{s}}$ at fixed total number of all other components of the system, in particular also of the electrons. Hence, we find that the GC energetics directly assesses the proton electrochemical potential  $\tilde{\mu}^{\mathrm{s}}_{{\mathrm{H^+}}}$ on the surface at applied potential conditions where
\begin{eqnarray}
\tilde{\mu}^{\mathrm{s}}_{{\mathrm{H^+}}} (\tilde{\mu}_{e^-}) \vcentcolon= \left.\frac{\partial \mathcal{G}^{\mathrm{s}}}{\partial  N_{\mathrm{H^+}}}\right|_{\tilde{\mu}_{e^-}} = \left.\frac{\partial G^{\mathrm{s}}}{\partial  N_{\mathrm{H^+}}}\right|_{N_{e^-}(\tilde{\mu}_{e^-})} \ . \label{eq:defsurfaceprotonlevel}
\end{eqnarray}
Hence eq.~(\ref{eq:GCdeltaG}) becomes
\begin{eqnarray}
\Delta \mathcal{G}=\tilde{\mu}^{\mathrm{s}}_{{\mathrm{H^+}}} - \tilde{\mu}_{{\mathrm{H^+}}}
\end{eqnarray}
where $\Delta \mathcal{G}$ simply and exclusively describes the difference of the proton electrochemical potential on the surface and in the solution (cf. Ref. \citenum{goodpaster2016identification}). In contrast to the canonical case, cf. eq. (\ref{eq:Can_DeltaG}), the electronic degrees of freedom only play a bystander role as they are fully considered and integrated into the electrode potential dependence of $\tilde{\mu}^{\mathrm{s}}_{{\mathrm{H^+}}}$ by means of the Legendre transform. These facts imply on the one hand that $\Delta \mathcal{G}$ remains meaningful independent on the charge state of the surface and simply describes the proton energetics equally in a chemisorbed state, in interfacial water or along an electrochemical reaction path. In contrast, the imposed charge balance in reaction equation (\ref{eq:reactionHads}) that underlies the simulated, simultaneous exchange of protons and electrons within the CC CHE method, makes the association of $\Delta G$ with the proton electrochemical potential non-obvious. Nonetheless, our rigorous finding that $\Delta \mathcal{G} = \Delta G$ proves the correctness of CC CHE calculations that assess such states (with partial electron transfer\cite{Chen2018Understanding}) by infinite cell size extrapolations of canonical CHE calculations at constant double layer charge $N^{\mathrm{net}}_{e^-}$.\cite{Chan2015Electrochemical}

\begin{figure}[t]
    \centering
    \includegraphics[width=0.6\textwidth]{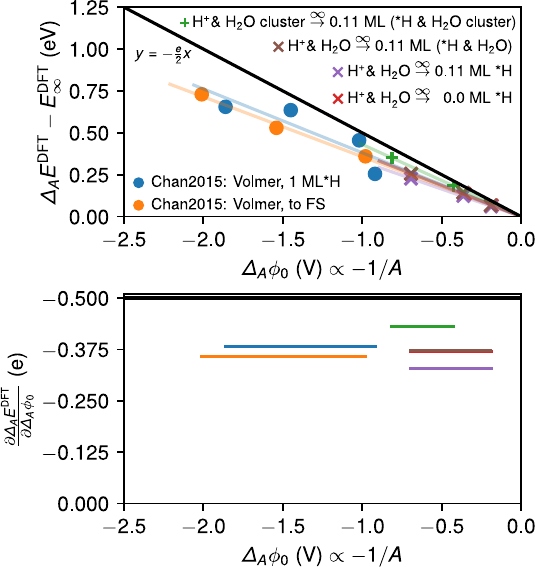}
    \caption{Observed cell size dependence for the energy difference $\Delta_{A} E^{\mathrm{DFT}}$ between a desorbed and an adsorbed proton. Datapoints taken from the literature (dots, Chan2015\cite{Chan2015Electrochemical}) describe the interface solvated by a static, full water layer. Results from own calculations in mixed implicit-explicit solvent environments (crosses and stars) employ only a single, static water molecule (crosses) or a static water cluster (stars), respectively. The lower panel reports the averaged size dependence as derived from linear fits through the data in the upper panel. See section S3 in the SI for more information.}
    \label{fig:Convergence_models}
\end{figure}

\paragraph{Size Convergence}
Such cell size extrapolations of canonical calculations are always necessary, when the considered adsorbate or adsorbate configuration induces a significant dipole in the system, as the native, canonical DFT energy obtained in finite-size periodic boundary condition  (PBC) supercells then exhibits significant, unphysical interactions with the periodic images. The significant magnitude of these finite-size errors in practically employed supercells was first recognized by analyzing the energy difference $\Delta_{A} E^{\mathrm{DFT}}= E^{\mathrm{DFT}}(\mathrm{H^+=des}) - E^{\mathrm{DFT}}(\mathrm{H^+=ads})$ between a desorbed and an adsorbed proton, as relevant in the study of proton adsorption in canonical, CC studies in a finite simulation cell size of area $A$.\cite{rossmeisl2008modeling,Skulason2010Modeling,Tripkovic2011Standard,Chan2015Electrochemical, gauthier2019practical} 
Along such a reaction path a dramatic change in the system's work function is observed that is associated with a change $\Delta_{A} \phi_0$ of the potential drop across the inner double layer. It was found\cite{rossmeisl2008modeling,Skulason2010Modeling,Tripkovic2011Standard} that $\Delta_{A} E^{\mathrm{DFT}}$ exhibits a significant cell size dependence with $\Delta_{A} E^{\mathrm{DFT}} \sim \Delta_{\infty} E^{\mathrm{DFT}} -0.5e \Delta_{A} \phi_0$ where $\Delta_{\infty} E^{\mathrm{DFT}}$ describes the energy difference between the desorbed and adsorbed state in the infinite cell size limit and $-0.5e \Delta_{A} \phi_0$ the approximate cell size  dependence of the energy difference. The latter term is explained by the energetics of (dis)charging a plate capacitor, created by the proton's charge when in the solvent and the according counter electron on the surface which can be considered as a (partial) charging of the double layer by $\Delta N^{\mathrm{net}}_{e^-} \sim 1$. For a metallic electrode of area $A$ and an inner layer capacitance $C$ this partial charging of the capacitor originates in the potential change $\Delta_{A} \phi_0 = -e/(CA)$ when the proton is in the desorbed state and the size-dependent scaling of $\Delta_{A} E^{\mathrm{DFT}}$ is thus explained by the change in the capacitively stored energy which amounts to 
\begin{eqnarray}
\Delta_{A} E^{\mathrm{cap}}(\Delta N^{\mathrm{net}}_{e^-}=1) &=& \frac{AC}{2} (\Delta_{A} \phi_0)^2 \label{eq:scalingRossmeisl0} \\
&=& -\frac{e}{2}\Delta_{A} \phi_0 \ . \label{eq:scalingRossmeisl}
\end{eqnarray}
While these insights and results were obtained based on an original internal energy referencing scheme\cite{rossmeisl2008modeling,Skulason2010Modeling,Tripkovic2011Standard}, the size convergence of standard, canonical CC energy differences replicate the (approximately) linear dependence of $\Delta_{A} E^{\mathrm{DFT}}$ on $\Delta_{A} \phi_0$ though with the magnitude of the prefactor being slightly smaller ($\sim 0.3e-0.4e$\cite{Chan2015Electrochemical}, see Fig.~\ref{fig:Convergence_models}). 

The capacitor expression, eq.~(\ref{eq:scalingRossmeisl}), not only provides a basis for extrapolating $\Delta_{A} E^{\mathrm{DFT}}$ to the infinite cell size limit $\Delta_{\infty} E^{\mathrm{DFT}}$ where $\Delta_{A} \phi_0\overset{A\rightarrow\infty}{=}0$, it also allows to estimate the considerable magnitude of finite size effects in CC calculations: For a prototypical, e.g. $(3 \times 3)$-fcc(111), supercell size with lattice constant $L_0 \sim 9\, \textnormal{\AA}$ ($A_0 \sim L_0^2 = 81 \textnormal{\AA}^2$) and with $C \sim 25\, \mu$F/cm$^2$ = 16 me/V$\textnormal{\AA}^2$\cite{rossmeisl2008modeling} one obtains $\vert\Delta_{A_0} \phi_0\vert = e/(A_0C) \sim 0.8$\,V and thus $\Delta_{A_0} E^{\mathrm{DFT}}-\Delta_{\infty} E^{\mathrm{DFT}} \sim 0.4$ eV for the finite-size error. This is in very good agreement with reported results obtained in calculations with protons in static, explicit water in contact with metallic electrodes and obviously far from negligible (cf. Fig.~\ref{fig:Convergence_models} and Refs. \citenum{rossmeisl2008modeling,Chan2015Electrochemical}).

It is important to note here, that while $\Delta_{A} E^{\mathrm{DFT}}$ in the respective studies refers to the energy difference between a desorbed and an adsorbed proton, the finite-size scaling is largely related only to the artificial PBC interactions of the desorbed proton state with its strong dipole.\cite{Chan2015Electrochemical, Abidi2022How} The adsorbed proton state typically only induces a minor work function change\cite{Hoermann2020Electrosorption,Surendralal2021Impact} and thus exhibits intrinsically only small PBC interactions.

In constant-potential simulations, any such finite-size errors are absent, as here any adsorbate dipole is by construction identically canceled by the counter-dipole induced by the adaption of electronic surface and electrolyte double-layer charges ($N_{e^-}^{\mathrm{net}}$) in order to keep the electrode potential constant. As a result, the constant-potential boundary conditions naturally compute the energetics at $\Delta_{A} \phi_0 =0$, and hence the PBC error $\Delta_{A} E^{\mathrm{cap}}=0$ at any cell size. Any remaining size effects are only due to higher-order multipole terms.

This fact can be seen most clearly by considering that the GC energy difference $\Delta_A \mathcal{E}^\mathrm{DFT}$ between a desorbed and an adsorbed proton in cell size $A$ in a constant-potential setup is given in lowest orders in the potential by the canonical energy difference $\Delta_A \left.E^{\mathrm{DFT}}\right|_{\mathrm{PZC}}$ at the PZC plus the energy change $\Delta_A G^{\mathrm{s}}_{\mathrm{exc,DL}}(U)$ due to adaption of the double-layer charge. As derived in Ref. \citenum{Hoermann2020Electrosorption}
\footnote{Cf. eq.~(14) in Ref. \citenum{Hoermann2020Electrosorption},
where states $\alpha$ and $\beta$ there are associated with the desorbed and the adsorbed proton state, here. In principle, $U$ in eq.~\ref{eq:estDL2} is the applied potential relative to the PZC of the state $\beta$, i.e., the slab with adsorbed proton. However, $U$ corresponds approximately to the applied potential relative to the PZC of the clean slab as the adsorbed proton has no (significant) dipole.\cite{Hoermann2020Electrosorption} As a result, we directly associate $U$ with the applied potential relative to the PZC of the clean slab for reasons of consistency with the consecutive discussions.}
\begin{eqnarray}
\Delta_A G^{\mathrm{s}}_{\mathrm{exc,DL}}(U) &=& A C U \cdot \Delta_A \phi_0 - \frac{A C}{2} (\Delta_A \phi_0)^2 + \dots \label{eq:estDL2}  \ , 
\end{eqnarray}
with $U=-\frac{1}{e}\left(\tilde{\mu}_{e^-}-\tilde{\mu}^{\mathrm{PZC}}_{e^-}\right) $ the applied potential relative to the PZC of the clean slab. While the first term in eq.~(\ref{eq:estDL2}) represents a dipole-field-like interaction term\cite{Hoermann2020Electrosorption,beinlich2022field,dominguezflores2023approximating} for the desorbed proton which is largely size invariant ($A\Delta_A \phi_0 \sim \textnormal{const.}$, see also section S1 in the SI and the discussion below), the second term in eq.~(\ref{eq:estDL2}) identically cancels the PBC error term in eq.~(\ref{eq:scalingRossmeisl0}) that is naturally present in the canonical PZC energetics $\Delta_A \left.E^{\mathrm{DFT}}\right|_{\mathrm{PZC}}$. These results underline that constant-potential calculations are intrinsically free of the significant, capacitor-like PBC interaction errors that are present in canonical CC calculations.

As a result, while the partial differential in eq.~(\ref{eq:GCdeltaG}) can be directly evaluated by finite differences of GC energies in moderate cell sizes ($L \sim 10$\,\AA), more care has to be taken when approximating the derivative in eq.~(\ref{eq:Can_DeltaG}) from finite differences of CC energies at such cell sizes.

\paragraph{Sensitivity to Modelling Errors}
While the discussion up to now only revealed disadvantages of canonical CC setups for studying the energetics at electrified metal-water interfaces, they do as well exhibit a variety of non-obvious advantages. There is first, of course, their more straightforward implementation in actual simulations, e.g. molecular dynamics setups. Instead of setting the potential $\tilde{\mu}_{e^-}$ explicitly, which necessitates (the non-trivial) implementation of a potentiostat and concomitant mitigation of existing convergence issues\cite{deissenbeck2021dielectric,deissenbeck2023dielectric, xia2023grand}, canonical CC calculations only necessitate a method to vary the surface charges $N^{\mathrm{net}}_{e^-}$. This can be achieved either using a continuum electrolyte model to screen the electronic excess charges on the surface\cite{Bonnet2013First,Ringe2019Understanding,Gauthier2019Challenges,Ringe2022Implicit} or by explicitly introducing real or artificial ions\cite{Skulason2007Density, rossmeisl2008modeling, Dudzinski2023First}. The induced energy change in such CC calculations can be described in general by a dipole-field-like energy contribution\cite{Koper2000Field,Liu2003Modeling,Noerskov2004Origin,Bonnet2013First,mccrum2015electrochemical,Ringe2019Understanding,Gauthier2019Unified,Hoermann2020Electrosorption,beinlich2022field,dominguezflores2023approximating,Dudzinski2023First} according to 
\begin{eqnarray}
\Delta_{A} G^{\mathrm{s}}_{\mathrm{DF}}(N^{\mathrm{net}}_{e^-}) &\approx& -\mathbf{E}(N^{\mathrm{net}}_{e^-}) \mathbf{D} \, \nonumber \\
&=& -eN^{\mathrm{net}}_{e^-} \Delta_A \phi_0 \quad , \label{eq:DF_energy} 
\end{eqnarray}
where $\mathbf{E}=-\frac{eN^{\mathrm{net}}_{e^-}}{\epsilon_0 A}$ is the average interfacial field and $\mathbf{D}= -\epsilon_0 A \Delta_A \phi_0 $ the adsorbate-induced surface dipole as deduced from the observed potential change $\Delta_A \phi_0$ in a cell of area $A$ (see also section S1 in the SI). In case of an identical charge vs. potential relation as in the GC calculation with $-eN^{\mathrm{net}}_{e^-}=A C U$, one obtains $\Delta_{A} G^{\mathrm{s}}_{\mathrm{DF}}=A C U \Delta_A \phi_0$ which is cell-size-independent and equivalent to the first term in eq.~(\ref{eq:estDL2}). These equivalent first order potential dependencies reflect nothing but the above prove of the equivalence of CC CHE and GC energetics in case that the charge vs. potential relations are identical and in case that PBC errors are negligible -- aka in the infinite cell size limit. 

Note, however, that the CC energetics in eq.~(\ref{eq:DF_energy}) depends only on $N^{\mathrm{net}}_{e^-}$ in agreement with a dipole-field energy that only depends on the interfacial field ($\propto N^{\mathrm{net}}_{e^-}/A$) and the adsorbate dipole ($\propto A \Delta_A \phi_0 $). In other words, $\Delta_{A} G^{\mathrm{s}}_{\mathrm{DF}}$ does not explicitly depend on $U$ or $C$ -- in contrast to eq.~(\ref{eq:estDL2}) -- and is thus rather insensitive to the counter charging method employed and the simulation-intrinsic capacitance $C$. It is thus largely irrelevant how the variation in $N^{\mathrm{net}}_{e^-}$ (or $\mathbf{E}$) is achieved in practice, which makes CC energy differences robust against variations in the counter-charge model and errors in the simulation-intrinsic interfacial capacitance, as demonstrated by
Gauthier \emph{et al.} in their assessment of the energetics of the proton adsorption problem\cite{Gauthier2019Unified}. Here, consistent energetics was observed when evaluating CC energy differences as a function of $N^{\mathrm{net}}_{e^-}$ but not when evaluating the energetics as a function of the system-intrinsic electrode potential $U$. Evidently, this robustness of CC energy differences to approximate $
\left.\frac{\mathrm{d} G^{\mathrm{s}}}{\mathrm{d}  N_{\mathrm{H}}}\right|_{N^{\mathrm{net}}_{e^-}}$ in eq.~(\ref{eq:Can_DeltaG}) derives from the significant energy (and thus error) cancellation when subtracting canonical energies $G^{\mathrm{s}}$ for $N^{\mathrm{net}}_{e^-} = \mathrm{const.}$ and thus in an environment of identical electrolyte composition. 

This fact has important implications, in particular, as the simulation of realistic capacitance curves $C(U)$ and hence $N^{\mathrm{net}}_{e^-}(U)$ is difficult in contemporary atomistic simulations: Typical implicit-model-derived capacitances are only approximately correct\cite{Nattino2018Continuum,Hoermann2019Grand,Schwarz2020electrochemical}, static interfacial water models miss the appropriate, dynamic solvent sampling and all-explicit, {\em ab initio} simulations are only possible for very high electrolyte concentrations (typically of the order of 1 M $\sim$ 1 ion pair per 55 water molecules). 

With $\Delta_{A} G^{\mathrm{s}}_{\mathrm{exc,DL}}(U) \propto A C U \cdot \Delta_{A} \phi_0$ depending explicitly on $U$ and $C$ in GC setups, any imperfect replication of the true charge vs. potential relation leads to errors in the GC energetics. In contrast, such errors can be (largely) mitigated by evaluating CC calculations not for the simulation-intrinsic $N^{\mathrm{net}}_{e^-}(U)$ relationship, but for an alternative, more realistic $N^{\mathrm{net}}_{e^-}(U)$ relation. This is possible as the electrode potential enters the CC energetics not at the electronic structure calculation level but only as a post-processing step when transforming the dipole-field energy, eq.~(\ref{eq:DF_energy}), into a function of $U$, i.e., via $-eN^{\mathrm{net}}_{e^-}(U) = ACU$. Similarly, one can derive from one calculation at given $N^{\mathrm{net}}_{e^-}$, the energetics for a variety of alternative interfacial conditions -- e.g. for different ion concentrations or ion types -- by assessing their impact on $C$ and using an appropriately altered $-eN^{\mathrm{net}}_{e^-}(U) = ACU$ relationship as e.g. done in Refs. \citenum{Ringe2019Understanding,Hoermann2021Thermodynamic,Hoermann2021Thermodynamica}. 

It is important to understand that using $N^{\mathrm{net}}_{e^-}(U)$ relationships that are different from the simulation-intrinsic ones is not merely an uncontrolled approximation but essentially encodes nothing but the principle of locality: While the potential $U$ and the capacitance $C$ are global properties that certainly affect the energetics as a function of $U$, the local energetics of adsorbates are affected (largely) by the local interactions. These are in good approximation given by the adsorbate-induced surface dipole $\mathbf{D}$ and the effective electrostatic field $\mathbf{E}$ in the vicinity of the surface\cite{beinlich2022field} which is essentially given by the electronic excess charge $N^{\mathrm{net}}_{e^-}$ on the surface. While the structure of the diffuse double layer certainly affects $C$ and hence the constant-potential energetics, the principle of locality suggests that it should not affect the local energetics. 

In summary, canonical CC calculations are natively less sensitive to the specifics of the double layer structure than GC calculations and are hence less prone to the errors introduced by their approximate description in actual computations. As a result, the CC approach allows to explicitly compute the stability of adsorbates under practically convenient conditions (e.g. high electrolyte concentration) and then to infer their potential dependence for other conditions in a post-processing step, by simply assuming an altered $N^{\mathrm{net}}_{e^-}(U)$ relationship\cite{Ringe2019Understanding}. 

Nonetheless, it is important to keep in mind, that changes in the electrolyte might not only alter the overall charge vs. potential relation. They could for instance also lead to more specific interactions of adsorbates with (partially solvated) electrolyte ions \cite{McCrum2016pH,Resasco2017Promoter,Kristoffersen2020Energy}, which would then not be captured by a simple rescaling of $N^{\mathrm{net}}_{e^-}(U)$ in a post-processing analysis. As a final remark, while yielding correct answers for the differential energetics, any energetic change from the self-consistent adaption of interfacial fields due to adsorbate-induced work function changes\cite{Huang2022Surface} is evidently also not included in the CC energetics whenever generic charging relations $N^{\mathrm{net}}_{e^-}(U)$ relative to the PZC of the clean surface are used. In contrast, these effects are included naturally in GC evaluations of the interfacial energetics at finite coverage and potential\cite{Hoermann2019Grand, Hoermann2021Thermodynamic} which is in particular relevant for understanding the (non-differential) interface energetics at high coverages with adsorbates that strongly affect the PZC.

\section{Conclusions}

Canonical CC and GC constant-potential calculations have emerged as the two prevalent approaches for first-principles atomic-scale simulations of metal electrodes in electrochemistry and -catalysis. Primarily due to practitioner-level approximations diverging results led initially to a separation into two schools, each with considerable scepticism about the other approach. With time, this situation improved, with general considerations on the convergence of different ensembles in the thermodynamic limit, empirical calculations\cite{Gauthier2019Unified} and low-order Taylor-expansions of canonical and GC energetics around the PZC\cite{Hoermann2020Electrosorption,beinlich2022field,dominguezflores2023approximating,beinlich2023controlled} providing increasing evidence that consistent results can be expected for the energetics of single adsorbates within CC and constant-potential calculations. The rigorous proof provided here finally provides a solid basis for a differentiated assessment of the equivalence and equal validity of both methods.

While fundamentally providing equivalent results in general, a remaining notable difference is the possibility of observing states in CC calculations that are unstable in the GC ensemble\cite{beinlich2023theoretical}. On a practical level, CC simulations are likely more robust against the specific description of the electrified interface and more straightforward and easier to implement, e.g. in molecular dynamics setups\cite{bonnet2012first, Bouzid2017Redox, surendralal2018first, deissenbeck2021dielectric, deissenbeck2023dielectric}. Nonetheless, they are more prone to finite-size errors than GC simulations in contemporary supercell setups. The latter point needs to be carefully addressed to obtain consistent differential energetics where $\Delta G = \Delta \mathcal{G}$ holds.

\begin{suppinfo}

This manuscript is accompanied by Supporting Information document which contains additional mathematical details as well as a detailed description of the data displayed in Fig.~1. The data and results from own calculations and a python script to reproduce Fig.~1 are available at \href{https://doi.org/10.17617/3.6W6BGK}{https://doi.org/10.17617/3.6W6BGK}.

\end{suppinfo}


\begin{acknowledgement}

The authors acknowledge funding and support from the German Research Foundation (DFG) under Germany's Excellence Strategy - EXC 2089/1- 390776260 (e-conversion), DFG project RE1509/33-1, as well as through the TUM International Graduate School of Science and Engineering (IGSSE).
All computations were performed on the HPC system Raven at the Max Planck Computing and Data Facility, which we gratefully acknowledge.

\end{acknowledgement}



\bibliography{main}

\end{document}